\documentclass[aps,twocolumn]{revtex4}
\usepackage{epsf}
 
\begin{document}
      \title{The flow equation approach to 
             the pairing instability problem}

     \author{T.\ Doma\'nski, A.\ Donabidowicz}
\affiliation{
             Institute of Physics, 
	     M.\ Curie Sk\l odowska University, 
             20-031 Lublin, Poland} 
      \date{\today}

\begin{abstract}
By means of the continuous unitary transformation similar to 
a general scheme of the Renormalization Group (RG) procedure 
we study the issue of symmetry breaking and pairing instability 
in the system of interacting fermions. Constructing a generalized 
version of the Bogoliubov transformation we show that formation 
of the fermion pairs and their superconductivity/superfluidity 
can appear at different temperatures. It is shown that strong 
quantum fluctuations can destroy the long-range order without 
breaking the fermion pairs which may still exist as incoherent 
and/or damped entities. Such unusual phase is characterized 
by a partial suppression of the density of states near 
the Fermi energy and by residual collective features like 
the sound-wave mode in the fermion pair spectrum. 
\end{abstract}

\maketitle


Formation of the fermion pairs is a common phenomenon for 
various physical systems of the interacting particles such as: 
electrons, nucleons, atoms and quarks.  Binding energy and 
the spatial extent of fermion pairs may vary from 
case to case depending on particular species and on specific 
interaction mechanism. To give some examples let us mention 
that pairing can be driven by: 
\begin{itemize}
\item[{(i)}]
exchange of phonons 
     (in classical superconductors, MgB$_{2}$, etc),
\item[{(ii)}]
exchange of magnons
     (superconductivity of the heavy fermion compounds), 
\item[{(iii)}]
 strong correlations
     (the high $T_{c}$ superconductors),
\item[{(iv)}]
Feshbach resonance
     (superfluidity of the ultracold fermion atoms),
\item[{(v)}]
or by other effects
      (nucleon pairing in nuclei, superfuidity of the neutron stars).
\end{itemize}

Usually formation of the fermion pairs goes hand in hand 
with appearance of the order parameter which consequently 
leads either to superconductivity (for charged particles such 
the conduction band electrons or holes) or to superfluidity 
(for electrically neutral objects like $^{3}$He or the ultracold 
fermion atoms in magnetooptical traps). However, a simultaneous
formation of pairs and emergence of the symmetry broken
phases  needs not be a rule. We will show here example 
that both these phenomena are distinct and happen to coincide 
at the same critical temperature $T_{c}$ only when the quantum 
fluctuations are weak.

\section{Hamiltonian of the interacting fermions} 

System of the interacting fermions can be described by 
the following Hamiltonian
\begin{eqnarray}
\hat{H} &=& \sum_{{\bf k},\sigma} (\epsilon_{\bf k}-\mu)
\hat{c}_{{\bf k}\sigma}^{\dagger}\hat{c}_{{\bf k}\sigma}
\label{general} \\
&+& \frac{1}{2}\sum_{{\bf k},{\bf k}',{\bf q}} 
\sum_{\sigma,\sigma'} U_{{\bf k},{\bf k}'}({\bf q})
\hat{c}_{{\bf k}+\frac{\bf q}{2},\sigma}^{\dagger}
\hat{c}_{{\bf k}'-\frac{\bf q}{2},\sigma'}^{\dagger}
\hat{c}_{{\bf k}'+\frac{\bf q}{2},\sigma'}
\hat{c}_{{\bf k}-\frac{\bf q}{2},\sigma}
\nonumber
\end{eqnarray}
where $\varepsilon_{\bf k}$ is a single particle energy 
for a given momentum ${\bf k}$ and $\sigma$ corresponds 
to additional quantum numbers like for instance spin 
$\uparrow$, $\downarrow$ for electrons, the angular 
momentum for atoms or the isospin for nucleons. The 
two-body interactions are described by the second term
with the potential $U_{{\bf k},{\bf k}'}({\bf q})$.
We use in (\ref{general}) the standard notation for 
the creation (annihilation) operators 
$\hat{c}_{{\bf k}\sigma}^{\dagger}$ 
($\hat{c}_{{\bf k}\sigma}$).

In general there can arise various kinds of ordering, 
for instance: ferromagnetism, antiferromagnetism, charge 
ordering, superconducting BCS state, etc. We will focus 
here on the pairing instabilities. For this purpose we 
further consider the Hamiltonian reduced only  to 
${\bf q}\!=\!{\bf 0}$ channel
\begin{eqnarray}
\hat{H} = \sum_{{\bf k},\sigma} \xi_{\bf k} 
\hat{c}_{{\bf k}\sigma}^{\dagger}\hat{c}_{{\bf k}\sigma} 
\;+\; \sum_{{\bf k},{\bf k}'} V_{{\bf k},{\bf k}'} \; \;
\hat{c}_{{\bf k}\uparrow}^{\dagger} \! \hat{c}_{-{\bf k}
\downarrow}^{\dagger} \; \hat{c}_{-{\bf k}'\downarrow} 
\hat{c}_{{\bf k}'\uparrow}
\label{BCS_hamil}
\end{eqnarray}
with $\xi_{\bf k}=\varepsilon_{\bf k}\!-\!\mu$ and we 
assume the two-body potential to be attractive $V_{{\bf k},
{\bf k}'}<0$. We will investigate this reduced BCS 
Hamiltonian using the nonperturbative method (described 
in section III) which belong to a family of the Renormalization 
Group techniques \cite{Wilson_NRG}.

\section{Renormalization Group approach} 

Thermodynamics of the system (total energy, specific heat, 
pressure, etc) can be computed from the partition function 
${\cal{Z}}=\mbox{Tr}e^{-\hat{H}/k_{B}T}$. It is convenient 
to express ${\cal{Z}}$ using the path integrals over the 
Grassmann variables $\psi_{{\bf k},\sigma}$, $\psi_{{\bf k}
,\sigma}^{*}$ (which formally represent the eigenvalues 
of the annihilation $\hat{c}_{{\bf k},\sigma}$ and 
creation $\hat{c}_{{\bf k},\sigma}^{\dagger}$ operators)
\begin{eqnarray}
{\cal{Z}}  =  \int D [ \psi,\psi^{*} ] \;\; e^{-S} .
\end{eqnarray}
The action consists of two contributions $S=S_{0}+S_{I}$ 
where the quadratic term
\begin{eqnarray}
S_0 = \sum_{\sigma} \int_{k} \psi^{*}_{{\bf k},\sigma} 
\; (i\omega_{n}\!-\!\xi_{\bf k} ) \; \psi_{{\bf k},\sigma}
\label{eqn4}
\end{eqnarray}
corresponds to a free part and integration in the equation
(\ref{eqn4}) runs over the four-vector $k \equiv (i\omega_{n},{\bf k})$ 
with the fermion Matsubara frequencies $\omega_{n}=(2n+1)
\pi k_{B}T$. The second quartic term refers to the two-body 
interactions
\begin{eqnarray}
S_I = - \int_{k,k'} V_{{\bf k},{\bf k}'}
\psi^{*}_{{\bf k},\uparrow} 
\psi^{*}_{-{\bf k},\downarrow}
\psi_{-{\bf k}',\downarrow}
\psi_{{\bf k}'\uparrow}
\end{eqnarray}
As far as the dynamic quantities are concerned (various
correlation functions) they are derivable directly from 
the generating functional
\begin{eqnarray}
&&{\cal{G}}[\chi,\chi^{*}]  = \label{eqn6} \\ &&
 \mbox{log} \left[ {\cal{Z}}^{-1}
\sum_{\sigma} \int D [ \psi,\psi^{*} ] e^{-(S + \int_{k}
\psi_{{\bf k},\sigma}^{*} \chi_{{\bf k},\sigma} + 
\psi_{{\bf k},\sigma} \chi_{{\bf k},\sigma}^{*})} \right]
\nonumber 
\end{eqnarray}
where $\chi_{{\bf k},\sigma}$ and $\chi_{{\bf k},\sigma}^{*}$ 
are the Grassman source fields. For instance, the single 
particle excitations can be determined via the two-point 
Green's function $\frac{\delta}{\delta \chi_{{\bf k},\sigma}
^{*}}\; \frac{\delta}{\delta \chi_{{\bf k},\sigma}} \; 
 {\cal{G}}[\chi,\chi^{*}]_{_{\chi=0,\chi^{*}=0}}$.

Physical properties of the system under consideration depend
predominantly on such fermion states which are located near 
the Fermi surface. Other states distant from the Fermi energy
are less relevant therefore it is useful to make a distinction 
between their contributions to the partition function
\begin{eqnarray}
{\cal{Z}} =   \int D^{<\Lambda}[\psi,\psi^{*}] 
\;  \int D^{>\Lambda}[\psi,\psi^{*}]
\;\; e^{S[\psi,\psi^{*}] } 
\end{eqnarray}
where symbol $D^{<\Lambda} [\psi,\psi^{*}]$ corresponds to
the fermion states whose distance from the Fermi energy is
smaller than a given cut-off $|\varepsilon_{\bf k}-
\varepsilon_{{\bf k}_{F}}| < \Lambda$. In the numerical RG
method \cite{Wilson_NRG} one first integrates out the high
energy fermion states. After completing such integration
one is left with only the low energy states. Partition
function is then ${\cal{Z}} =   \int D^{<\Lambda}
[\psi,\psi^{*}] \;\; e^{-S^{\Lambda}[\psi,\psi^{*}] }$
where the renormalized action is defined by 
\begin{eqnarray}
e^{-S^{\Lambda}[\psi,\psi^{*}]} = \int D^{>\Lambda}
[\psi,\psi^{*}] \;\; e^{-S[\psi,\psi^{*}]} .
\label{eqn8}
\end{eqnarray}
This action (\ref{eqn8}) can be then cast into (\ref{eqn6})
simplifying the calculation of the generating functional 
\begin{eqnarray}
&& {\cal{G}}[\chi,\chi^{*}]   = \\ 
& & \mbox{log} \left[ {\cal{Z}}^{-1}
\int D^{<\Lambda} [ \psi,\psi^{*} ] 
e^{-(S^{\Lambda} + \int_{k}^{<\Lambda} \psi_{\bf k}^{*} 
\chi_{\bf k} + \psi_{\bf k} \chi_{\bf k}^{*})} .
\right] 
\nonumber 
\end{eqnarray}
Such idea of {\em mode elimination} has been introduced in 
the theoretical physics a long time ago \cite{scaling_idea}. 
Wilson adopted it to the solid state physics by proposing 
a sequential integration of the fermion fields down to 
some small cut-off $\Lambda$. Reducing bit by bit the 
cut-off $\Lambda$ to infinitesimally small values he was
able to study emergence of the critical phenomena 
\cite{Wilson_NRG}. For a more specific discussion 
of various RG formulations we recommend the the review 
papers \cite{RG_review}. 

However, in the case of symmetry broken phases (such as
the superconducting state) the simple scaling procedure 
usually fails due to a natural lower boundary cut-off
(the energy gap $\Delta$ of the single particle excitations).
It has been even claimed that conventional RG techniques 
are blind with respect to the symmetry-broken states 
\cite{Gersch-06}. The situation is not that much hopeless,
there are possible routes to circumvent this problem. 
Let us mention two of them
\begin{itemize}
\item[{(i)}]
one can impose by hand an infinitesimal symmetry-breaking 
parameter $\Delta(\Lambda)$ at a certain initial condition 
$\Lambda\!=\!\Lambda_{0}$ and then its physical meaning 
would eventually establish upon the flow $\Lambda 
\rightarrow 0$ \cite{Salmhofer-04},
\item[{(ii)}] 
in order to eliminate the interacting part $S_{I}$ one can
enlarge the Hilbert space by introducing the linear coupling 
to the boson fields $\Phi$, $\Phi^{*}$ via the Hubbard 
- Stratonovich transformation \cite{Kopietz-05}.
\end{itemize}  

Using the first procedure one must be cautious how to
constrain the small symmetry breaking parameter while 
in the latter method there rise additional complications
dealing with the new boson fields. Perhaps the second
option is more natural because after all the interactions
are always mediated by certain boson fields (phonons, 
photons, gravitons, etc). Unfortunately in practice 
it is hard to go beyond a simple saddle point 
approximation except than by taking into account 
the small Gaussian corrections around it. In the next 
section we shall briefly describe a more convenient 
procedure which has been proposed in 1994.

\section{Continuous canonical transformation} 

An alternative approach to deal with the many-body effects 
(which in particular is suitable for studying the symmetry
broken phases) has been invented by Wegner \cite{Wegner-94} 
and independently by Wilson and G\l azek \cite{Wilson-94}.
Instead of integrating out the high energy states (fast 
modes) one constructs a continuous unitary transformation
$\hat{H}(l) = \hat{U}(l) \hat{H} \hat{U}^{\dagger}(l)$ 
with a purpose to simplify the Hamiltonian either to diagonal
or at least to a block-diagonal form. This is achieved through 
a series of infinitesimal steps $\hat{H} \longrightarrow 
\hat{H}(l) \longrightarrow \hat{H}(\infty)$, where $l$ 
stands for some formal continuous parameter. In a course 
of transformation the Hamiltonian evolves according to
the differential {\em flow equation}
\begin{eqnarray}
\partial_{l} \hat{H}(l) = \left[ \hat{\eta}(l), 
\hat{H}(l) \right]
\end{eqnarray}
where the generating operator is defined as $\hat{\eta}(l) 
= - \hat{U}(l) \; \partial_{l} \hat{U}^{\dagger}(l)$.

This method has a similarity to the traditional RG 
scaling procedure because
\begin{itemize}
\item[{(i)}]
     diagonalization of the high energy sector occurs 
     mainly during initial part
\item[{(ii)}]
     while the low energy states are worked out at 
     the very end of transformation.
\end{itemize}
Roughly speaking, a role of the Wilson's cut-off 
energy $\Lambda$ is played here by $\Lambda^{l} = 
\frac{1}{\sqrt{l}}$.

Advantage of such new procedure becomes particularly clear 
when investigating the mutual relations between the high 
and small energy states (i.e.\ between the {\em fast} and 
{\em small} modes). In the continuous canonical transformation 
one treats both energy sectors on equal footing throughout 
the whole transformation process. Thereof their feedback 
effects are feasible to study.

For the Hamiltonians $\hat{H}=\hat{H}_{0}+\hat{H}_{1}$ 
(where $H_{0}$ is a diagonal part in a given representation) 
it has been proposed \cite{Wegner-94} to choose 
\begin{eqnarray}
\hat{\eta}(l) = \left[ \hat{H}_{0}(l), 
\hat{H}_{1}(l) \right]
\label{eta}
\end{eqnarray}
so that the off-diagonal term is eliminated
at the asymptotic point
\begin{eqnarray}
\lim_{l \rightarrow \infty}\hat{H}_{1}(l)=0 .
\end{eqnarray}
There are also other possible ways for constructing 
the generating operator $\hat{\eta}$. Some a survey 
we recommend the recent review papers 
\cite{Wegner-06,Kehrein-06}.

As far as the diagonalization is concerned the continuous
canonical transformation turns out to be a rather convenient
tool. However, if one needs the correlation functions then 
the situation becomes more cumbersome. For example, to 
determine the correlation functions $\langle \hat{A}(t) 
\hat{B}(t') \rangle$ one needs the statistical average 
\begin{eqnarray}
\mbox{Tr} \left\{  e^{-\beta \hat{H}} \hat{O} \right\} & = & 
\mbox{Tr} \left\{  \hat{U}(l)  e^{-\beta \hat{H}} 
\hat{O} \hat{U}^{\dagger}(l) \right\}  \nonumber \\
& = & 
\mbox{Tr} \left\{  e^{-\beta {\hat{H}(l)}} 
\hat{O}(l) \right\}  
\label{averaging} 
\end{eqnarray}
where ${\hat{O}(l)}=\hat{U}(l) {\hat{O}} \hat{U}^{\dagger}(l)$. 
The easiest way to carry out the statistical averaging 
(\ref{averaging}) is in the limit $l \longrightarrow \infty$ 
when $\hat{H}(\infty)$ becomes (block-)diagonal. However, 
this requires a simultaneous transformation of the observables 
$\hat{O} \longrightarrow  \hat{O}(l) \longrightarrow \hat{O}
(\infty)$. The corresponding flow equation is given in 
a familiar form
\begin{eqnarray}
\partial_{l} \hat{O}(l) = \left[ \hat{\eta}(l), 
\hat{O}(l) \right].
\label{O_flow}
\end{eqnarray}
Thus, calculation of the correlation functions is rather 
more relative to the projection techniques.

\section{The bilinear Hamiltonian} 

To illustrate how the flow equation method actually works 
we first consider the bilinear Hamiltonian
\begin{eqnarray}
\hat{H} & = & \sum_{{\bf k},\sigma} \xi_{\bf k} 
\hat{c}_{{\bf k}\sigma}^{\dagger} \hat{c}_{{\bf k}\sigma} 
- \sum_{\bf k} \left( \Delta_{\bf k} \; \hat{c}_{{\bf k}
\uparrow}^{\dagger} \hat{c}_{-{\bf k}\downarrow}^{\dagger} 
+ \Delta^{*}_{\bf k} \; \hat{c}_{-{\bf k}\downarrow}
\hat{c}_{{\bf k} \uparrow} \right) 
\label{hamil}
\end{eqnarray}
which can be solved exactly, in particular by a single step 
Bogoliubov transformation \cite{Bogoliubov}. The off-diagonal 
terms in the Hamiltonian (\ref{hamil}) can be thought as 
resulting from the mean field approximation for the weak 
pairing potential $V_{{\bf k},{\bf k}'}<0$ with a usual
definition of the order parameter $\Delta_{\bf k}
=-\sum_{{\bf k}'} V_{{\bf k},{\bf k}'} \langle 
\hat{c}_{-{\bf k}'\downarrow} \hat{c}_{{\bf k}'\uparrow}
\rangle$. 

To get the rigorous solution we now construct a continuous 
transformation such that the convoluted states $|{\bf k},
\uparrow>$ and $|-{\bf k},\downarrow>$ will be decoupled. 
This continuous process depends on a distance from 
the Fermi surface $|\xi_{\bf k}|$ (and the same holds 
for computation of the coherence factors $u_{\bf k}$, 
$v_{\bf k}$). 

Using the Wegner's proposal (\ref{eta}) we obtain the 
two coupled flow equations
\begin{eqnarray}
\partial_{l} \; \xi_{\bf k}(l) & = & 4 \xi_{\bf k}(l) 
|\Delta_{\bf k}(l)|^{2} \label{xi_flow} \\
\partial_{l} \; \Delta_{\bf k}(l)  & = & -4 (\xi_{\bf k}(l))^{2}
\Delta_{\bf k}^{*}(l)  \label{Delta_flow} 
\end{eqnarray} 
for $l$-dependent quantities $\xi_{\bf k}(l)$ and 
$\Delta_{\bf k}(l)$. The second equation (\ref{Delta_flow})
yields
\begin{eqnarray} 
| \Delta_{\bf k}(l)| = |\Delta_{\bf k}| \mbox{exp} 
\left\{ -4\int_{0}^{l} dl' [\xi_{\bf k}(l')]^2 \right\}
\end{eqnarray}
which proves that indeed the off-diagonal terms gradually 
diminish under the flow $l \rightarrow \infty$. There is 
a singular point ${\bf k}\!=\!{\bf k}_F$ which is unaffected 
by the transformation but in the thermodynamic limit (i.e.\ 
for a macroscopic number of particles $N$) its role 
becomes marginal.

By combining the equations (\ref{xi_flow},\ref{Delta_flow}) 
one gets the following invariance
\begin{eqnarray} 
\partial_{l} \left\{ (\xi_{\bf k}(l))^{2} + 
|\Delta_{\bf k}(l)|^{2} \right\} = 0 .
\label{invar} 
\end{eqnarray} 
which implies that in the limit $l\rightarrow\infty$ the 
eigenvalues have a character of the Bogoliubov spectrum
\cite{Domanski-06,Stelter}
\begin{eqnarray}
\xi_{\bf k}(\infty) = \mbox{sgn}({\xi_{\bf k}}) 
\sqrt{(\xi_{\bf k})^{2} + |\Delta_{\bf k}|^{2} } .
\label{bogol}
\end{eqnarray} 
%
\begin{figure}
{\epsfxsize=5.5cm{\epsffile{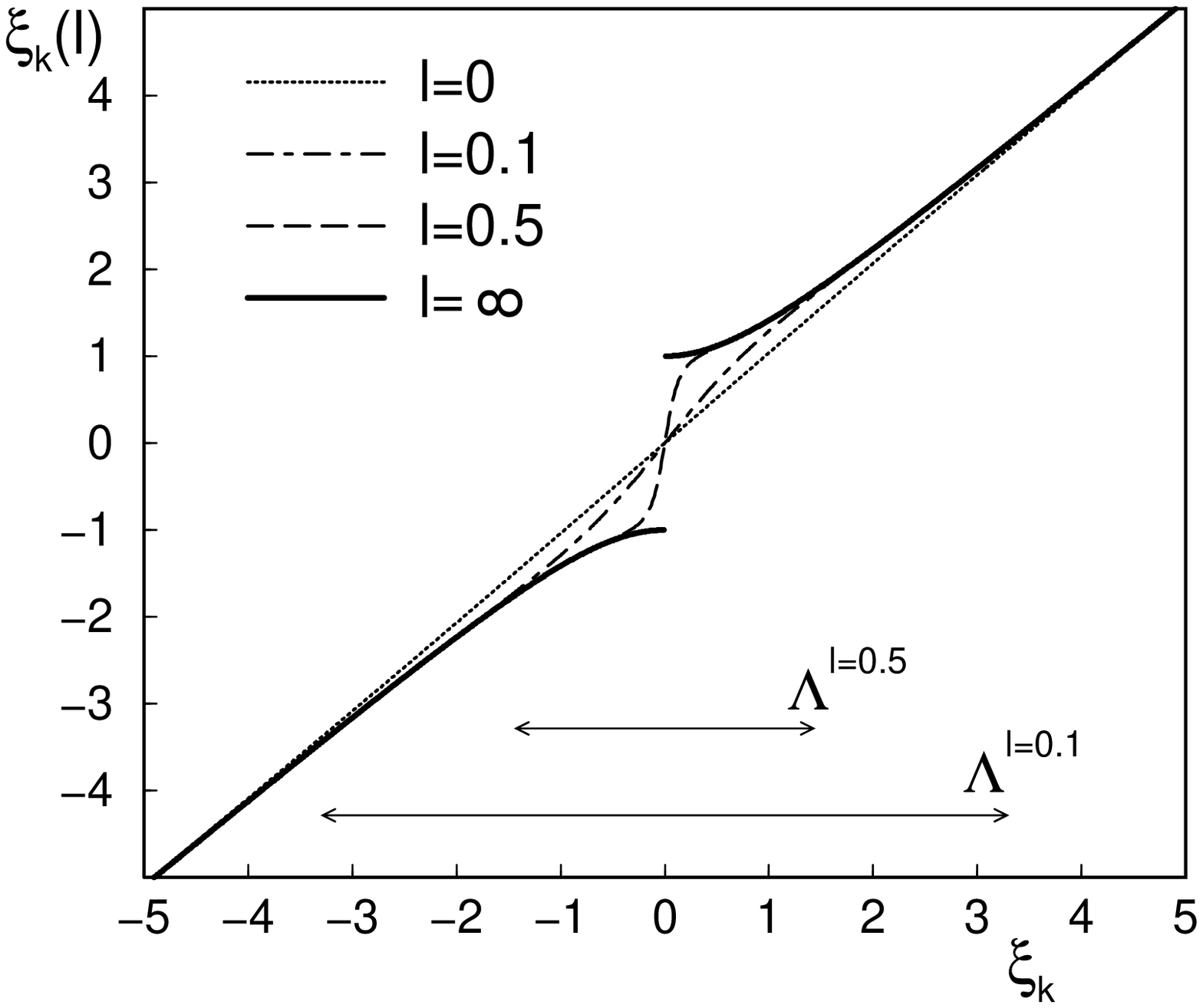}}} \hspace{0.5cm}
{\epsfxsize=5.5cm{\epsffile{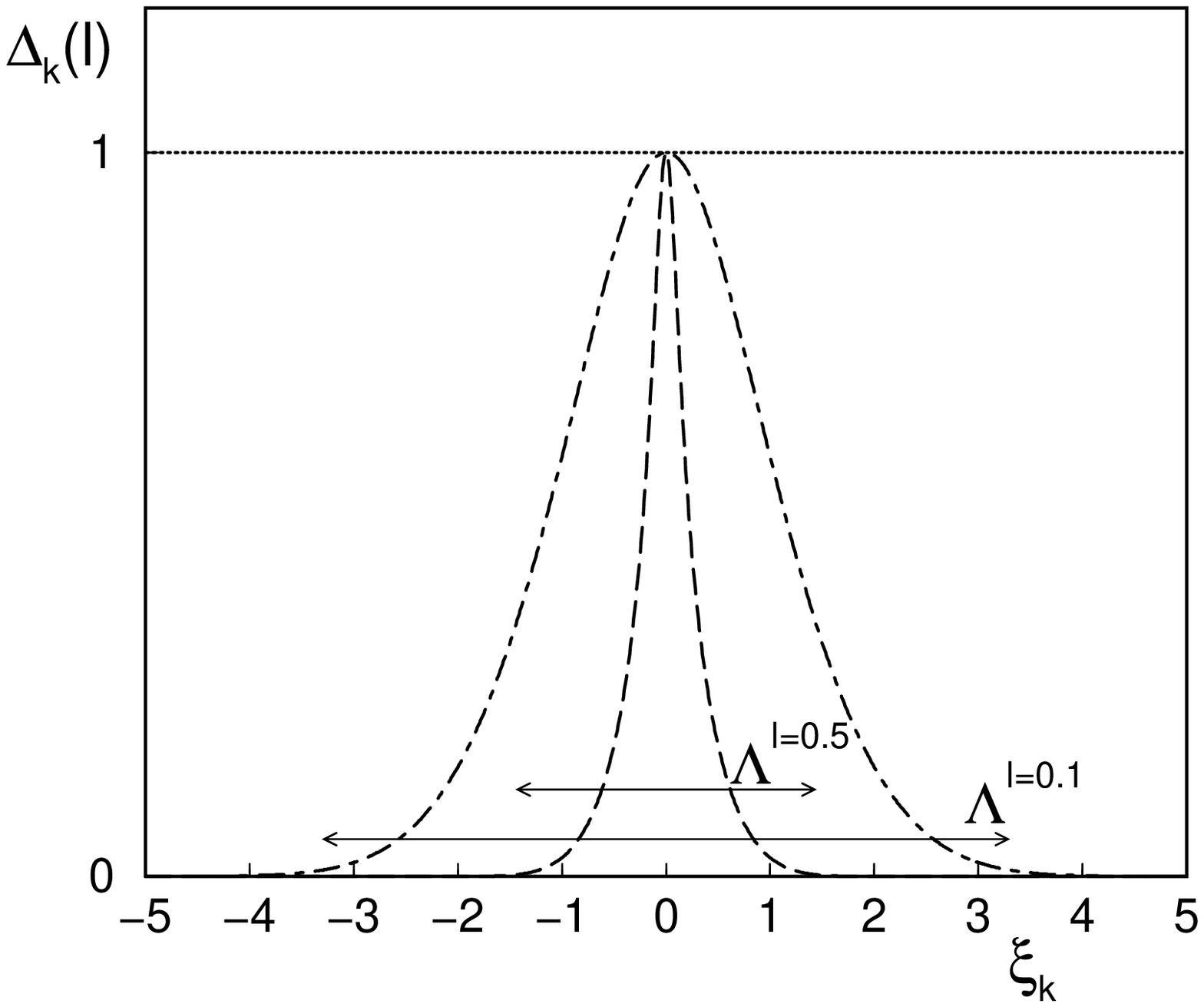}}}
\caption{Variation of $\Delta_{\bf k}(l)$ (panel on the right)
and $\xi_{\bf k}(l)$ (panel on the left) versus a distance 
from the Fermi surface $\xi_{{\bf k}_{F}}=0$. We assumed 
a constant parameter $\Delta_{\bf k}(0)=\Delta$ in the model
Hamiltonian (\ref{hamil}). Energies presented in this figure 
are expressed in units
of $\Delta$ and the flow parameter in units of $\Delta^{-2}$.
Note that for a given $l$ the renormalizations are practically
completed outside the energy window of the width $\Lambda^{l}=
\frac{1}{\sqrt{l}}$ as marked by the arrows.}
\label{Fig1}
\end{figure}

In figure \ref{Fig1} we illustrate evolution of the parameter 
$\Delta_{\bf k}(l)$ which initially was assumed to be constant 
$\Delta_{\bf k}=\Delta$. In the first turn disappearance 
of $\Delta_{\bf k}(l)$ occurs for states distant from the 
Fermi energy and then in the second turn to states located 
nearby to $\mu$. This evolution is accompanied by the 
renormalization of fermion energies $\xi_{\bf k}(l)$. Finally 
for $l \rightarrow \infty$ the quasiparticle energies evolve 
to the gaped Bogoliubov dispersion (\ref{bogol}).

In order to specify the complete single particle spectrum
we now derive the correlation function  $\langle \hat{c}
_{{\bf k}\sigma} (t) \hat{c}_{{\bf k}\sigma}^{\dagger}
(t')\rangle$ where time evolution is given by the standard 
relation $\hat{O}(t)=e^{it\hat{H}}\hat{O}e^{-it\hat{H}}$.
As emphasized in the previous section this requires 
determination of the $l$-dependent single particle operators.
From a detailed analysis \cite{Domanski-06} we obtain 
the Bogoliubov-type parameterizations 
\begin{eqnarray}
\hat{c}_{{\bf k}\uparrow}(l) & = & u_{\bf k}(l) \hat{c}_{{\bf k}
\uparrow}+v_{\bf k}(l) \hat{c}_{-{\bf k}\downarrow}^{\dagger}
\label{Ansatz1} \\
\hat{c}_{-{\bf k}\downarrow}(l)^{\dagger} & = & 
- v_{\bf k}(l) \hat{c}_{{\bf k}\uparrow} 
+ u_{\bf k}(l) \hat{c}_{-{\bf k}\downarrow}^{\dagger} 
\label{Ansatz2}
\end{eqnarray} 
with the initial values $u_{\bf k}(0)=1$, $v_{\bf k}(0)=0$.
From (\ref{O_flow}) we derive the following differential 
equations for the coefficients $u_{\bf k}(l)$, $v_{\bf k}(l)$
\begin{eqnarray}
\partial_{l}  \; u_{\bf k}(l) & = & 2 \xi_{\bf k}(l)
\Delta_{\bf k}(l) \; v_{\bf k}(l),
\label{u_flow} \\
\partial_{l}  \; v_{\bf k}(l) & = & -2 \xi_{\bf k}(l)
\Delta_{\bf k}(l) \; u_{\bf k}(l) .
\label{v_flow}
\end{eqnarray}  
It can be easily checked that equations (\ref{u_flow},
\ref{v_flow}) lead to the following invariance of $l$-dependent
coefficients $|u_{\bf k}(l)|^2  + |v_{\bf k}(l)|^2\!=\!1$. 
This invariance assures that the fermion anticommutation 
relations are obeyed for arbitrary level of the transformation
$\left\{ \hat{c}_{{\bf k}\sigma}(l),\hat{c}_{{\bf k'}
\sigma'}^{\dagger}(l)\right\} = \delta_{{\bf k},{\bf k'}} 
\delta_{\sigma,\sigma'}$. 

\begin{figure}
{\epsfxsize=5.5cm{\epsffile{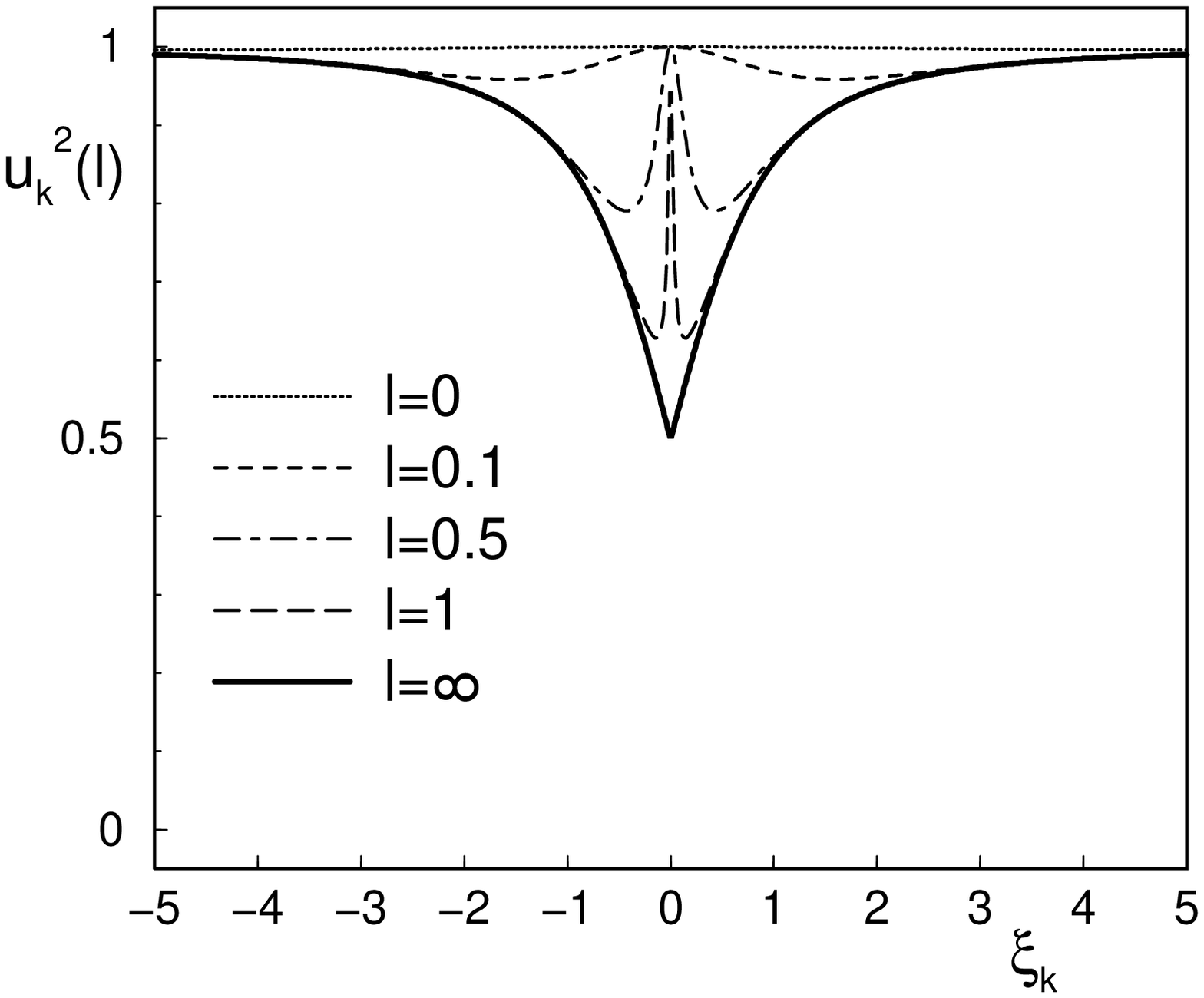}}} \hspace{0.5cm}
{\epsfxsize=5.5cm{\epsffile{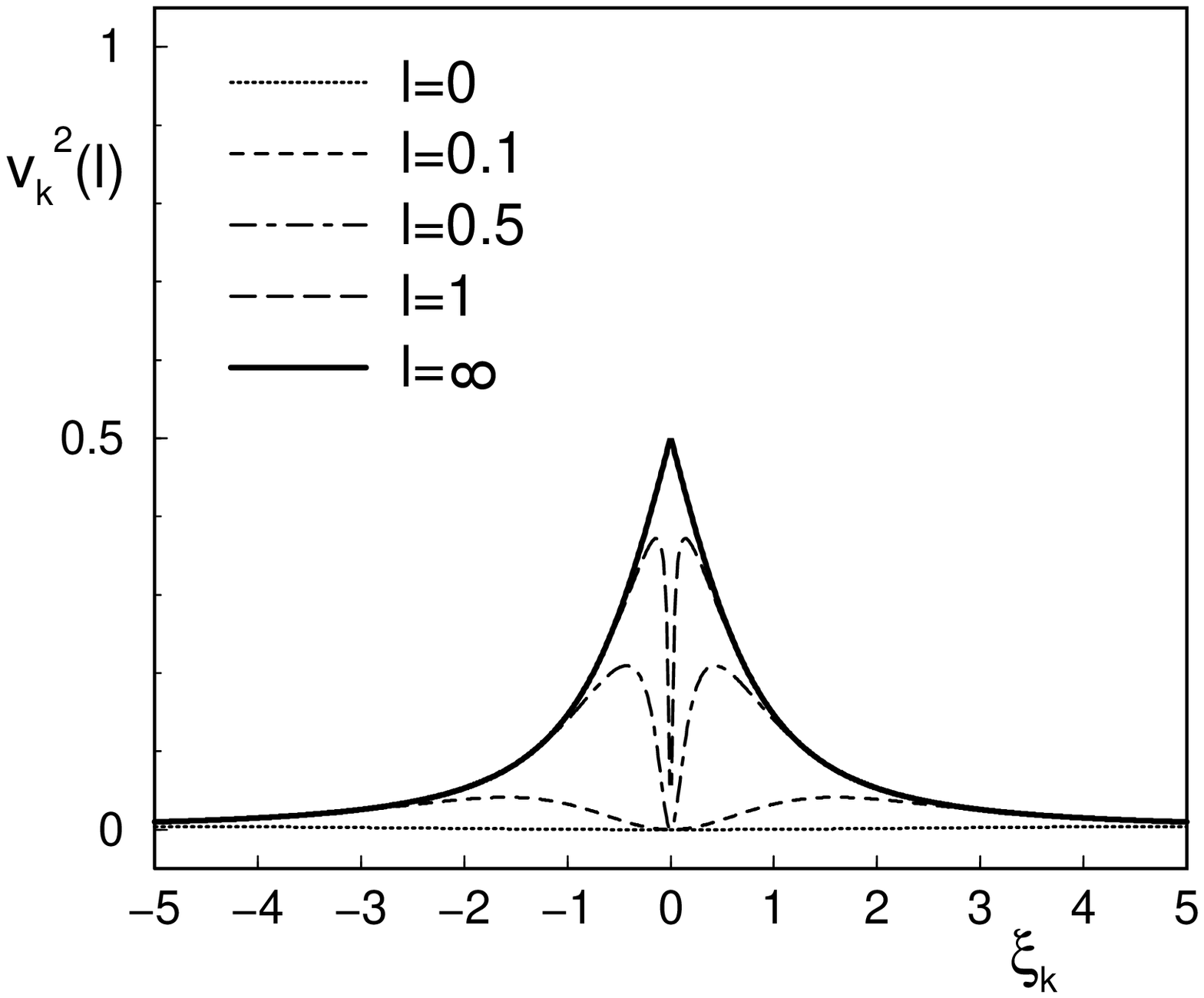}}}
\caption{The BCS coherence factors $u_{\bf k}^{2}(l)$ and
$u_{\bf k}^{2}(l)$ versus the energy $\xi_{\bf k}$ for some 
representative values of $l$ as indicated in the legend.}
\label{Fig2}
\end{figure}

We have previously shown \cite{Domanski-06} that the asymptotic 
$l\!=\!\infty$ values of the coefficients $u_{\bf k}(l)$ and 
$v_{\bf k}(l)$ coincide with the usual BCS factors
\begin{eqnarray} 
u_{\bf k}^{2}(\infty) 
& = & \frac{1}{2} \left[ 1 + \frac{\xi_{\bf k}}
{\sqrt{(\xi_{\bf k})^{2} + |\Delta_{\bf k}|^{2} } } 
\right] \\
u_{\bf k}(\infty)v_{\bf k}(\infty) & = &\frac{\Delta_{\bf k}}
{2\sqrt{(\xi_{\bf k})^{2} + |\Delta_{\bf k}|^{2} }} 
\end{eqnarray}
and $v_{\bf k}^{2}(\infty) = 1 -  u_{\bf k}^{2}(\infty)$.
In figure \ref{Fig2} we show the $l$-dependent factors 
$u_{\bf k}^{2}(l)$, $v_{\bf k}^{2}(l)$ versus the energy 
measured from the Fermi surface. We can notice that for 
arbitrary $l$ the asymptotic values are reached by all 
fermion states located outside the energy cut-off 
$\Lambda^{l} = \frac{1}{\sqrt{l}}$.

The complete single particle excitation spectrum is made 
of the particle and hole contributions. The corresponding
spectral function is found \cite{Domanski-06} to be 
\begin{eqnarray}
A({\bf k},\omega)  = u_{\bf k}^{2}(\infty)
\delta( \omega - \xi_{\bf k}(\infty) )
+ v_{\bf k}^{2}(\infty)
\delta( \omega + \xi_{\bf k}(\infty) )
\nonumber \\ \label{A_BCS}
\end{eqnarray} 
In a straightforward manner we can arrive at the following
expressions for the average occupancy of ${\bf k}$ momentum
and for the order parameter
\begin{eqnarray}
\langle  \hat{c}_{{\bf k}\uparrow}^{\dagger} 
\hat{c}_{{\bf k}\uparrow} \rangle & = & \frac{1}{2} 
\left[ 1 - \frac{\xi_{\bf k}}{\xi_{\bf k}(\infty)} 
\mbox{tanh}\frac{
\xi_{\bf k}(\infty)}{2k_{B}T} \right], 
\label{eq26}
\\
\langle \hat{c}_{-{\bf k}\downarrow} \hat{c}_{{\bf k}\uparrow} 
\rangle & = & \frac{\Delta_{\bf k}}{2\xi_{\bf k}(\infty)} 
\mbox{tanh}\frac{\xi_{\bf k}(\infty)}{2k_{B}T}.  
\label{eq27}
\end{eqnarray} 
The Bogoliubov spectrum (\ref{bogol}) together with the
expectation values expressed in equations (\ref{eq26},\ref{eq27}) 
reproduce the rigorous solution for the bilinear Hamiltonian 
(\ref{hamil}).

\section{Pairing in the strongly correlated fermion system} 

The two particle interactions in (\ref{general}) or in the reduced 
Hamiltonian (\ref{BCS_hamil}) can be canceled out exactly by
introducing the Hubbard-Stratonovich fields $\Phi$, $\Phi^{*}$.
Skipping a detailed derivation we assign here the following 
effective Hamiltonian to the resulting fermion and boson 
degrees of freedom \cite{BF}
\begin{eqnarray}
\hat{H} & = &    \sum_{{\bf k}\sigma} \xi_{\bf k} 
        \hat{c}_{{\bf k}\sigma}^{\dagger} \hat{c}_{{\bf k}\sigma}
        + \sum_{\bf q} \left( E_{\bf q} - 2 \mu \right) 
	\hat{b}_{\bf q}^{\dagger} b_{\bf q} 
\nonumber \\ &+& 
        \frac{1}{\sqrt{N}}  \sum_{{\bf k},{\bf q}} 
	v_{{\bf k},{\bf q}} \left[   
        \hat{b}_{\bf q}^{\dagger} \hat{c}_{-{\bf k},\downarrow} 
	\hat{c}_{{\bf k}+{\bf q},\uparrow} 
	+ \mbox{h.c.} \right]
\label{BF_hamil}
\end{eqnarray}
In the simplest step one can study (\ref{BF_hamil}) on the 
level of saddle point approximation when boson fields are replaced 
by constant number. This is a mean field approximation and
formally it is equivalent to linearization of the boson-fermion 
term via
\begin{eqnarray}
        \hat{b}_{\bf q}^{\dagger}  \hat{c}_{-{\bf k},\downarrow} 
	\hat{c}_{{\bf k}+{\bf q},\uparrow} & \simeq & 
	\langle \hat{b}_{\bf q}^{\dagger} \rangle 
	\hat{c}_{-{\bf k},\downarrow} \hat{c}_{{\bf k}+{\bf q},\uparrow}
	+ \hat{b}_{\bf q}^{\dagger} \langle \hat{c}_{-{\bf k},\downarrow} 
	\hat{c}_{{\bf k}+{\bf q},\uparrow}  \rangle
\nonumber \\ & - &
        \langle \hat{b}_{\bf q}^{\dagger} \rangle 
	\; \langle \hat{c}_{-{\bf k},\downarrow} 
	\hat{c}_{{\bf k}+{\bf q},\uparrow}  \rangle .
\label{meanfield_BF}
\end{eqnarray}
When substituting (\ref{meanfield_BF}) to the Hamiltonian 
(\ref{BF_hamil}) the fermion and boson subsystems  become 
decoupled from one another. The physical fermion part 
acquires then the bilinear structure (\ref{hamil}) with the parameter 
$\Delta_{\bf k} = v_{{\bf k},{\bf 0}} \langle \hat{b}_{{\bf q}
\!=\!{\bf 0}} \rangle $. The mean field properties of 
the model (\ref{BF_hamil})  have been summarized in 
the review paper \cite{Micnas-90}.

In this section we show how to go beyond the mean field 
scenario using the flow equation method. Our main strategy 
is to design a continuous canonical transformation $\hat{U}(l)$
which, step by step, dismantles the fermion from boson degrees 
of freedom. Technical remarks concerning such transformation has 
been given in our previous work \cite{Domanski-01} where we 
con\-strained the Hamiltonian into $\hat{H}=\hat{H}_{0}+
\hat{H}_{B-F}$ with $\hat{H}_{0}=\hat{H}_{F}+\hat{H}_{B}$ 
denoting the independent fermion and boson contributions 
and $\hat{H}_{B-F}$ corresponding to their interaction. We 
have followed the idea proposed by Wegner \cite{Wegner-94} by 
choosing $\hat{\eta}(l) = [\hat{H}_{0}(l),\hat{H}_{B-F}(l)]$
which yields
\begin{eqnarray}
\hat{\eta}(l) &=&  \frac{1}{\sqrt{N}} \sum_{{\bf k},{\bf q}}
\left[ E_{\bf q}(l) - \varepsilon_{-{\bf k}}(l) - 
\varepsilon_{{\bf k}+{\bf q}}(l) \right] 
\nonumber \\ & \times &
\left( v_{{\bf k},{\bf q}}(l)
 \hat{b}_{\bf q}^{\dagger} \hat{c}_{-{\bf k}\downarrow} 
\hat{c}_{{\bf k}+{\bf k} \uparrow} - \mbox{h.c.} \right) .
\label{BF_eta}
\end{eqnarray}

All the $l$-dependent quantities have been determined by us
selfconsistently using the iterative numerical Runge Kutta algorithm
to solve the set of differential flow equations (16-21) presented 
in the Ref.\ \cite{Domanski-01}. 
We studied a situation with the fixed number of particles 
in the system $n_{tot}=\sum_{{\bf k},\sigma} \langle 
\hat{c}_{{\bf k} \sigma}^{\dagger} \hat{c}_{{\bf k} \sigma} 
\rangle + 2 \sum_{\bf q} \langle \hat{b}_{\bf q}^{\dagger} 
\hat{b}_{\bf q} \rangle$. Among the important physical 
effects obtained for the fixed point $l \rightarrow \infty$ 
we could point out that:
\begin{itemize}
\item[{(c)}]
boson particles (which can be thought as the fermion pairs) 
acquire a finite mobility owing to the interaction with the 
itinerant fermions, 
\item[{(b)}] 
fermions in turn are affected by the boson particles and 
this effect shows up by a loss of the single particle states 
near the Fermi surface (such depletion of the density of states 
is often called in the literature as pseudogap),
\item[{(c)}]
in addition to the pseudogap there appears a resonant scattering 
between fermions \cite{Domanski-pra03}. 
\end{itemize}
Let us remark that the resonant-type scattering processes have 
been known for a long time in the nuclear physics. Such Feshbach 
mechanism is nowadays widely explored experimentally and 
theoretically for the atomic gasses cooled to ultralow 
temperatures enabling observation of such quantum phenomena 
like the Bose Einstein condensation (BEC) and ultimately 
also the atomic superfuidity.

\begin{figure}
\centerline{\epsfxsize=9cm{\epsffile{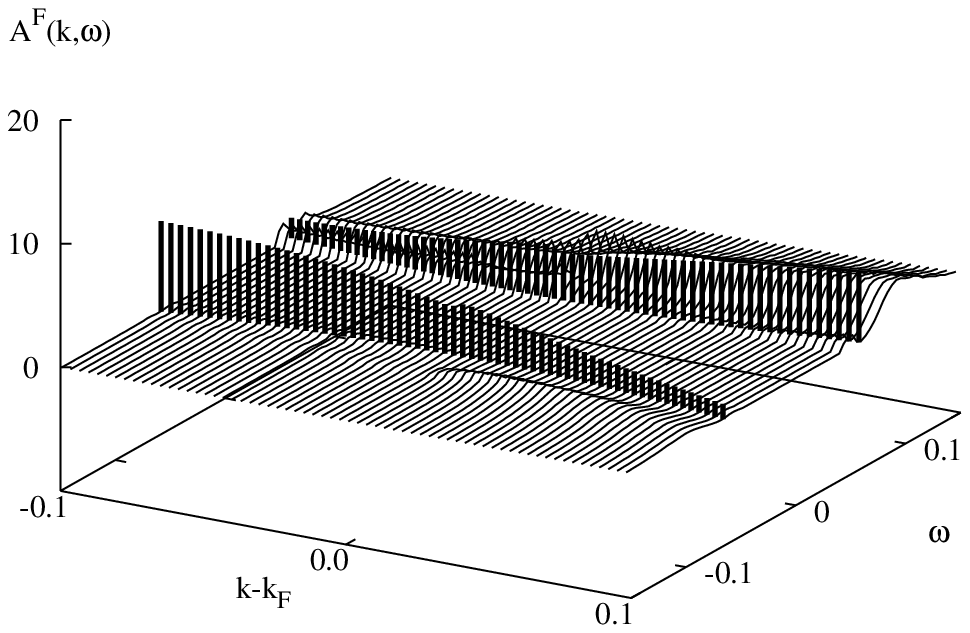}}}
\vspace{-1.5cm}
\centerline{\epsfxsize=9cm{\epsffile{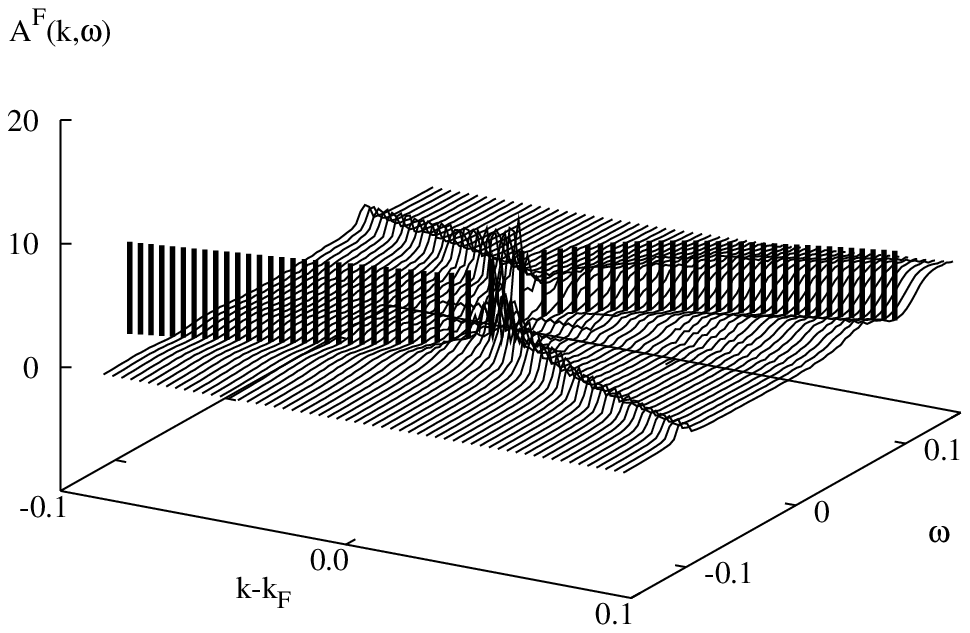}}}
\caption{The effective single particle spectrum of fermions
$A^{F}({\bf k},\omega)$ for the superconducting state (panel 
on the left h.s.) and in the normal state near the critical 
temperature $T > T_{c}$ (panel on the right h.s.).}
\label{Fig3}
\end{figure}

Since fermion states located near the Fermi energy get combined 
with the boson species it is natural to expect that the single
and two-particle properties are going to affect each other. We
studied systematically their interplay within the flow equation
method. To derive the single particle spectrum we had to transform
the annihilation $\hat{c}_{{\bf k}\sigma}$ and creation 
$\hat{c}_{{\bf k}\sigma}^{\dagger}$ operators using
(\ref{BF_eta}). From the flow equation (\ref{O_flow}) we
derived the following generalized Bogoliubov transformation
\cite{Domanski_prl}
\begin{eqnarray}
&&{\hat{c}_{{\bf k}\uparrow}(l) \choose \hat{c}_{-{\bf k}\downarrow}^{\dagger}(l)} 
 =   {u_{\bf k}(l) \choose -v_{\bf k}^{*}(l)} \;
\hat{c}_{{\bf k}\uparrow} + {v_{\bf k}(l) \choose u_{\bf k}^{*}(l)} \; 
\hat{c}_{-{\bf k}\downarrow}^{\dagger} \;\;\;\;\;\;\;\;  
 \nonumber \\
&  & + \; \frac{1}{\sqrt{N}} 
\sum_{{\bf q} \neq{\bf 0}} \left[
{p_{{\bf k},{\bf q}}(l) \choose r_{{\bf k},{\bf q}}^{*}(l)} \; 
\hat{b}_{\bf q}^{\dagger}
\hat{c}_{{\bf q}+{\bf k}\uparrow}  + 
{r_{{\bf k},{\bf q}}(l) \choose -p_{{\bf k},{\bf q}}^{*}(l)} \; \hat{b}_{\bf q} 
\hat{c}_{{\bf q}-{\bf k}\downarrow}^{\dagger}
\right] . 
\nonumber \\ & &
\label{BF_bogoliubov}
\end{eqnarray}
In consequence the single particle spectral function was found 
to have a different structure than the usual BCS result (\ref{A_BCS}). 
For temperatures $T<T_{c}$ it consists of two narrow quasiparticle 
peaks at energies $\omega = \pm \sqrt{(\varepsilon_{\bf k}-\mu)^2 
+ \Delta_{sc}^{2}}$ and a certain amount of the damped states 
forming an incoherent background outside the superconducting gap
(see the left panel in figure \ref{Fig3}). When traversing 
the critical temperature $T_c$ to a normal state the gaped 
Bogoliubov-type spectrum seems to be preserved, however for
increasing temperature the {\em shadow branch} becomes more and 
more damped (see the right h.s.\ panel in figure \ref{Fig3}).
Physically it means that fermion pairs have no longer 
an infinite life-time. Finally, for temperatures exceeding 
a certain characteristic value $T^{*}$ the Bogoliubov modes 
are completely gone and there remains only a single peak 
at the renormalized energy $\xi_{\bf k}(\infty)$ 
\cite{Domanski_prl}.

To check a direct impact of the above mentioned behavior 
on the pair correlations we investigated the following 
correlation function
\begin{eqnarray}
\sum_{{\bf k},{\bf k}'} \left< \; \hat{c}_{-{\bf k} 
\downarrow} (t) \hat{c}_{{\bf q}+{\bf k}\uparrow}(t) 
\;\; \hat{c}_{{\bf q}+{\bf k}'\uparrow}^{\dagger}(t') 
\hat{c}_{-{\bf k}' \downarrow}^{\dagger}(t') \right>  .
\end{eqnarray}
From the flow equation (\ref{O_flow}) for the pair 
operators $\sum_{\bf k} \hat{c}_{-{\bf k} \downarrow} 
\hat{c}_{{\bf q}+{\bf k}\uparrow}$ we obtained the 
corresponding spectral function
\begin{eqnarray}
A_{pair}({\bf q},\omega) = {\cal{N}}_{\bf q}  \;
\delta \left( \omega - \tilde{E}_{\bf q} \right)
 + {\cal{A}}_{\bf k}^{inc} \left( \omega \right) .
\end{eqnarray}
It contains the quasiparticle peak at $\omega=\tilde{E}_{\bf q}$
and the incoherent background ${\cal{A}}_{\bf k}^{inc}( \omega )$.
For $T<T_{c}$ the quasiparticle peak is well separated from 
the incoherent background and, in the limit ${\bf q} \rightarrow 
{\bf 0}$, has the important sound-wave dispersion 
$\tilde{E}_{\bf q} = c \; |{\bf q}|$ (like the collective 
sound-wave branch in the superfluid state of $^{4}$He). 
Unfortunately in the case of charged fermions such
as the conduction band electrons this mode is usually 
pushed to the huge plasma frequency because of the strong 
Coulomb repulsion. For the electrically neutral objects 
this {\em Goldstone mode} is a hallmark of the symmetry 
broken phase. 

\begin{figure}
\centerline{\epsfxsize=6.5cm{\epsffile{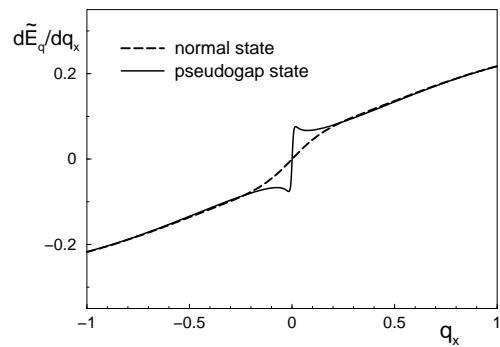}}}
\caption{Derivative of the fermion pair dispersion 
$d\tilde{E}_{\bf q}/d q_{x}$ (where $\tilde{E}_{\bf q} 
\equiv E_{\bf q}(\infty)$) for the normal (the dashed 
line) and for the pseudogap state close to $T_{c}$ 
(the solid curve). Note a tendency for a qualitative 
changeover from the parabolic to linear relation of
$E_{\bf q}(\infty)$.}
\label{Fig4}
\end{figure}

Above the transition temperature and close to $T_{c}$ 
there are still some residual collective features
possible to observe. At small momenta the qusiparticle 
peak $\omega\!=\!\tilde{E}_{\bf q}$ overlaps with the 
incoherent background therefore the long wavelength
limit is not suitable for appearance of the collective
effects (strictly speaking for ${\bf q} \rightarrow {\bf 0}$ 
the Goldstone mode is replaced by a parabolic dispersion).
A remnant of the Goldstone mode splits off from the incoherent
background for finite momenta above a certain critical value 
${\bf q}_{crit}$. This is illustrated in figure \ref{Fig4}.

Collective features seen in the pair-pair correlations
(or in the density-density correlations) tell us directly 
about a presence or absence of the long range order which 
is necessary for the onset superconductivity/superfluidity. 
As a matter of fact the single particle properties are then 
ill defined (at least in a vicinity of the Fermi energy). 
Instead of single fermions one should rather think about
fermion pairs as good quasiparticles. Although the fermion
pairs are necessary for superconudctivity/superfluidity
the other way around this is not valid. It has been shown 
by Eagles \cite{Eagles-69}, Nozieres and Schmitt-Rink 
\cite{Nozieres-85} and by a number of other authors 
that value of the binding energy (magnitude of the gap 
in the single particle spectrum) does not scale linearly
with the superfluid phase stiffness $n_{s}$ which 
determines the transition temperature $T_{c}$. Our work 
indeed confirms that existence of the preformed (i.e.\ 
incoherent) fermion pairs is a natural expectation while 
approaching $T_{c}$ from above. Such type of situation 
can be encountered in the high $T_c$ cuprate superconductors
where the strong quantum fluctuations are driven by the 
unscreened Coulomb repulsion between electrons and by 
the reduced dimensionality of CuO$_{2}$ planes \cite{PALee-06}. 
Pseudogap have been also unambiguously observed in the 
ultracold fermion atoms close to the unitarity limit 
(i.e.\ on the Feshbach resonance) \cite{Levin-05}.

\section{Summary} 

We have presented  the method of continuous unitary 
transformation originating from a general scheme of the 
RG scaling procedure. This non-perturbative technique 
overcomes usual problems of the standard RG methods 
in application to the symmetry broken states. We have
illustrated it on the exactly solvable case of the 
bilinear Hamiltonian. This new method is moreover 
capable to study the possible feedback effects between 
the {\em fast} and {\em slow} modes treating both 
sectors simultaneously throughout the whole continuous 
transformation.

Applying this method to the strongly interacting fermion system 
we have shown that formation of the fermion pairs need not 
be accompanied by the transition to superfluid/superconducting 
state. Strong quantum fluctuations can suppress the long-range 
coherence (ordering) so that effectively the fermion pairs 
exist even in the normal state above $T_{c}$.  Such preformed 
fermion pairs could be observed experimentally by e.g.\ probing 
the single particle spectrum (using the STM or ARPES spectroscopies) 
or in measurements of the correlations between pairs (via any 
experimental technique sensitive to the pair susceptibility).

\section*{Acknowledgements}

This work is partly supported by the KBN grant 
2P03B06225. We acknowledge fruitful discussions with 
J.\ Ranninger and K.I.\ Wysoki\'nski.

\end{document}